\documentclass[journal]{IEEEtran}
\ifCLASSINFOpdf
\else
\fi
\usepackage{mathrsfs}
\usepackage{amsthm}
\usepackage{makeidx}
\usepackage{examples}
\usepackage[pdftex]{graphicx}  
\usepackage{amssymb}
\usepackage{algorithm}
\usepackage{epstopdf}
\usepackage{color}
\usepackage{url}
\usepackage{caption}
\usepackage{algorithm}
\usepackage{algorithmic}
\usepackage{amsmath}
\usepackage{mathtools}

\usepackage{hyperref}
\usepackage[utf8]{inputenc}

\usepackage{lipsum}
\makeatletter
\def\footnoterule{\relax%
  \kern-5pt
  \hbox to \columnwidth{\hfill\vrule width 0.5\columnwidth height 0.4pt\hfill}
  \kern4.6pt}
\makeatother

\begin{document}

\title{BacSoft: A Tool to Archive Data on Bacteria$^{1}$\thanks{$^{1}$Poster abstract, DNA Computing and Molecular Programming $24^{th}$ International Conference, DNA 24, Jinan, China, October 8-12, 2018}}

\author{
\IEEEauthorblockN{Amay Agrawal\IEEEauthorrefmark{1}, Dixita Limbachiya\IEEEauthorrefmark{1}, Ravikumar M\IEEEauthorrefmark{2}, Taslimarif Saiyed \IEEEauthorrefmark{2} and Manish K. Gupta\IEEEauthorrefmark{1}\\}

\IEEEauthorblockA{\IEEEauthorrefmark{1}Laboratory of Natural Information Processing, Dhirubhai Ambani Institute of Information and Communication Technology, Gandhinagar, Gujarat, 382007 India. \\ Email: amayagrawal22@gmail.com, dlimbachiya@acm.org and mankg@computer.org}\\ 
\IEEEauthorblockA{\IEEEauthorrefmark{2}Centre for Cellular and Molecular Platforms,
GKVK Post, Bellary Road,
Bangalore 560065, India. \\ Email: mravikumar@ccamp.res.in, taslim@ccamp.res.in, 
}

}


\maketitle
\begin{abstract}
Recently, DNA data storage systems have attracted many researchers worldwide. Motivated by the success stories of such systems, in this work we propose a software called BacSoft to clone the data in a bacterial plasmid by using the concept of genetic engineering. We consider the encoding schemes such that it satisfies constraints significant for bacterial data storage. 
\end{abstract}


%
\IEEEpeerreviewmaketitle

\section{Introduction} 
The amount of digital data generated is increasing at really high speed, and it is estimated to be around  35 Zettabyte ($10^{21}$) bytes by 2020. This digital universe consisting of the big data needs a reliable, dense and affordable storage medium to store data. Recently, Biocomputers have attracted researchers from computer science and engineering to apply their principles to living cells \cite{green2017complex}. One of the applications is to use the nature hard drive as a storage medium which is DNA. DNA consists of four bases namely A (Adenine), T (Thymine), G (Guanine) and C (Cytosine) that encodes the information of life. DNA is the most reliable source of data storage that has evolved through generations therefore it is natural to use DNA for digital data storage.

In DNA data storage system, data is encoded into strings of A, T, G and C using different encoding schemes. These data encoded DNA is synthesized and stored in the appropriate environmental conditions. Stored data can be decoded back to the original data by using DNA sequencing. In last few years, researchers have developed DNA based data storage systems by introducing different encoding schemes which showcased that such systems are robust \cite{grass2015robust}, reliable \cite{bornholt2016dna} and dense \cite{yazdi2015dna}. Storing the data on DNA was first showcased by G. Church \textit{et al.} \cite{church2012next} and N. Goldman \textit{et al.} \cite{goldman2013towards} in the year 2012 and 2013 respectively. Motivated by this, a software DNACloud was developed by a team at Gupta Lab, that demonstrated the data encoding in DNA sequences using a modified Goldman's scheme  \cite{DNAcloud} . In subsequent years, various DNA based data storage systems were proposed that revealed the density and durability of the DNA for the archival data storage \cite{limbachiya2015natural}. A rewritable and random access DNA based data storage was proposed by J. Bornholt \textit{et al.} \cite{bornholt2016dna} and Yazdi \textit{et al.} \cite{yazdi2015rewritable}. Very recently, Y. Erlich and D.Zeilinski \cite{erlich2017dna}  proposed capacity achieving codes by developing fountain codes to encode the data in DNA.

M.Ailenberg \textit{et al.} and N.Yachie \textit{et al.} \cite{ailenberg2009improved}  \cite{yachie2007alignment}  pioneered the idea of using the bacteria for the data storage by giving the proof of bacterial data storage at a small scale. In Yachie \textit{et al.} method, they converted the famous Einstein relativity equation $E=MC^2$ in \textit{B.subtillius} soil bacteria to store information DNA in multiple loci of genomic DNA of bacteria repeatedly. They could successfully encode $120$ bits in $4.2$ MB genome of the bacteria and decoded it back by multiple sequence alignment (MSA). Another method based on PCR (Polymerase Chain Reaction) technique, was proposed by Bancroft \textit{et al.} \cite{bancroft2001long}. Recently, S.L Shipman \textit{et al.} at the Wyss Institute for Biologically Inspired Engineering and Harvard Medical School, used bacterial living cells, for the first time to store movie using a powerful gene editing tool CRISPR \cite{shipman2016molecular}. Very recently in \cite{tavella2018dna}, a systematic framework to simulate the data storage in a bacterial plasmid is given by considering the bacterial plasmids as clusters. To use bacteria for data storage, some of the DNA constraints important for DNA data storage \cite{limbachiya2018contraint} are common to bacterial data storage which is fixed GC content (DNA sequence with the fixed number of Gs and Cs) and no homopolymers (DNA sequence without repeated DNA bases, e.g. ATGCTG). In this work, we propose a method which generates DNA sequences with both these constraints.

All the previous work for bacteria based data storage systems have been performed under experimental trials without using any automated software. In order to ease the process of encoding the data into bacterial DNA, software is required that can verify the experimental protocols before hands-on experiments. To achieve this, motivated by the genetic engineering protocol, we introduce the software to encode the data in bacterial plasmids. Genetic engineering deals with manipulating the bacteria by the insertion of a foreign DNA into plasmids. The plasmid is a carrier that allows the insertion of an external DNA and transfers it to bacteria. 

In this work, we have built an open source software, which can be used to automatically store data in the bacterial plasmid using the concepts of genetic engineering. The software generates the DNA sequences with GC content (50 $\%$) and avoids long runs of DNA sequences (no homopolymer). First, we encode the text data into DNA sequences using the encoding strategy discussed in the section \ref{encoding}. After encoding, this data is cloned into the plasmid which can be selected from the list of different available plasmids. Restriction enzymes, which helps in inserting data into plasmid are automatically selected in the software based on the input DNA sequence. After Cloning, this data can be decloned back from plasmid and then decoded into the original data. A feature of Gel Electrophoresis simulation of encoded DNA is included in the software that allows visualizing the DNA on the simulated gel according to the length of DNA. All the steps described above can be visualized adequately in the software.

This paper is organized as follows. \textit{Section 2} discuss the algorithm used for Encoding the data, Cloning the data into a plasmid and then Decloning and Decoding the data back to original form. \textit{Section 3} discuss the GUI of our software. \textit{Section 4} contains the functionality and workflow of the software. \textit{Section 5} contains some examples showing how data can be encoded, cloned, decoded and decoded in the software. \textit{Section 6} provides the link to the website from where the software can be downloaded. \textit{Section 7} concludes the paper with some general remarks.   
 
\section{Algorithm}
This section describes algorithms used for encoding the text data into the bacterial plasmid. It includes Cloning the data into the plasmid, Decloning it back from plasmids and the Decoding it back to original data.

\subsection{Encoding the text data } \label{encoding}
To encode the data, first, a file is selected using the import function of the software.


After importing the file, it is encoded into the corresponding DNA sequences that will be cloned into the bacterial plasmid. To encode the data, we have used the following encoding scheme:

\begin{itemize}
    \item First, the corresponding text file is converted into the binary format using the standard ASCII conversion table.
    \item Then this binary file is divided into n $(x_{1}, x_{2}, x_{3},......,x_{n})$ chunks of equal length, where the length of each chunk is 32 bits.
    \item Next, we form k $(m_{1}, m_{2},m_{3},....,m_{k})$ new chunks out of the original n chunks using XOR operation (addition modulo $2$) on three consecutive chunks. 
    \item Now an 8-bit header is added to the above chunk, which is used for indexing. 
    So the total chunk size is 40 bits (32 bits data + 8 bits header).
    \item Next, encode the corresponding binary block into DNA by using the Table \ref{encodetable} which converts binary to DNA. In this, if the first bit of the binary block is '\textit{0}' then it is encoded as 'G' else if it is '1' then it is encoded as 'C'.
   

    \begin{table}[ht]
    \begin{center}
    \begin{tabular}{|c c c|} 
        \hline
        Previous bps & 0 & 1 \\ [0.5ex] 
         \hline\hline
         A & C & G  \\ 
         \hline
         T & G & C \\
         \hline
         G & A & T  \\
         \hline
         C & T & A \\
         \hline
        \end{tabular}
         \end{center}
         \caption{Conversion table from binary bits to DNA nucleotides}
         \label{encodetable}
        \end{table}
     \item The above encoding scheme is designed in such a way that the homopolymer runs and GC content are already taken care of while in the method proposed by \cite{erlich2017dna}, an additional step of screening is required for testing these DNA constraints.
\end{itemize}

\subsection{Cloning the Data}



\begin{figure*}
\centering
\includegraphics[scale=0.5]{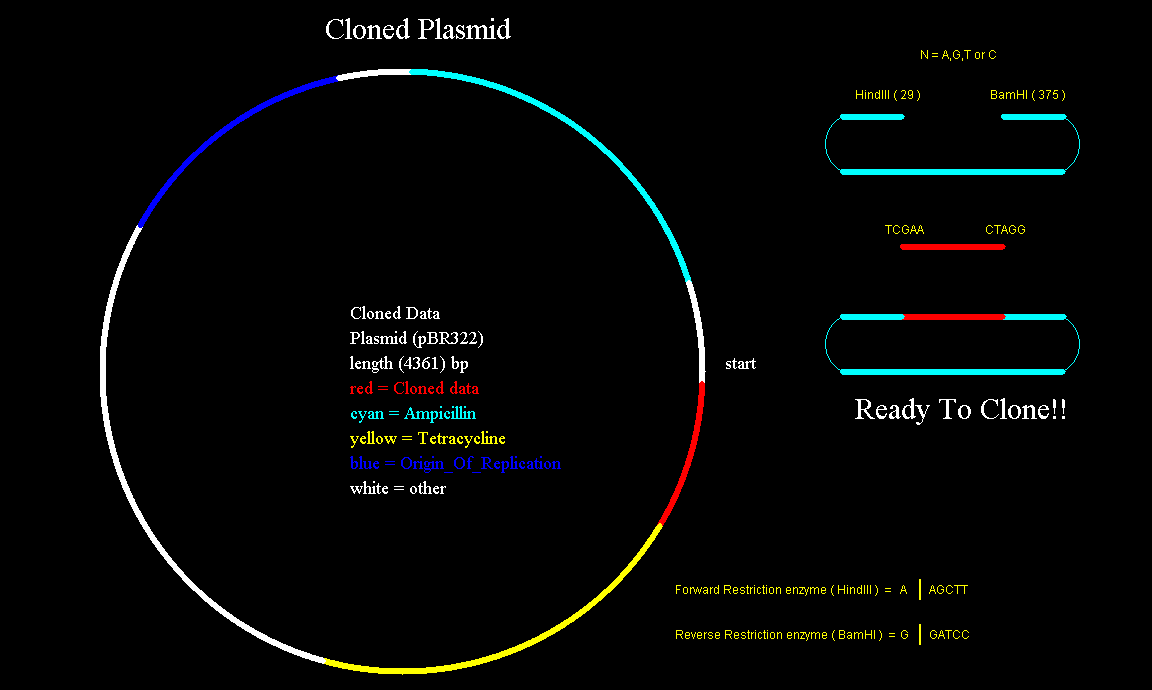}
\caption{Cloning process explained, with selected MCS and Restriction Enzyme used, and their corresponding sequence}
\label{capture3}
\end{figure*}

Once the data is encoded, the user can select any "Desirable plasmid" and the "Restriction Enzyme" category from the drop-down menu available. There are 176 plasmids available (e.g., pBR322, pUC18, pUC19 and others). There are five different restriction enzymes category available like 6$+$ Cutters, restriction\_enzymes, Unique and Dual\_Cutters, Unique 6$+$ Cutters, Unique Cutters. After selecting the plasmid, if the imported text data size exceeds the maximum limit of plasmid, then a pop-up warning will be displayed. At present, we have been successfully able to clone around 10 Kb of data into the plasmids. The plasmids with maximum insertion capacity are pJAZZ-OK and pJAZZ-OC belonging to $E.Coli$ bacteria. On selecting the "MCS and Restriction Enzyme" button, Multiple Cloning Sites (MCS) and the plasmid diagram appears on the screen. MCS diagram shown in the Figure \ref{capture}, describes the various restriction enzymes available in the corresponding plasmid along with their cloning sites. The plasmid diagram (see the Figure \ref{capture2}), describes various elements like antibiotic resistance markers (e.g.  Ampicillin, Tetracycline, Chloramphenicol, Kanamycin etc.), repressors, gene and other elements present in the corresponding plasmid. On selecting the "Clone Data" button, the data gets cloned into the selected plasmid and the Cloned Plasmid diagram appears as shown in the Figure \ref{capture3}. For Cloning purposes it employs the following steps: 

\begin{itemize}
    \item First, software scans for the Cloning site with sticky ends in the plasmid.
    \item Next, it checks whether the encoded data sequences at start contains complementary base pairs corresponding to the restriction enzyme. If not, then it inserts the corresponding complementary base pairs at the start of the encoded data.
    \item After determining the restriction site, it checks the encoded data length and finds another cloning site according to size.
    \item Once it identifies the cloning sites, it checks for the complementary base pairs at the end of the encoded data to stick into the plasmid. If it is not found, then it inserts the complementary base pairs corresponding to the restriction enzyme at the end of the encoded data.
    \item The data is cloned into the plasmid.
\end{itemize}

The Cloning process described above can be seen in the Cloned Plasmid diagram as shown in Figure \ref{capture3}. In the figure \ref{capture3}, one can see three diagrams on the top right corner which are explained as follows:

\begin{itemize}
    \item First diagram shows both the cloning sites of plasmid along with the restriction enzymes that were selected for cloning the data. Restriction sites of the displayed restriction enzymes are shown at the bottom of the screen.
    \item Second diagram shows the DNA sequence at both the ends of our encoded data.
    \item Third diagram shows that the DNA sequences of encoded data are complementary to restriction sites of selected restriction enzymes at both ends. 
    Therefore the data gets successfully inserted into the plasmid.
\end{itemize}

\subsection{Decloning and Decoding Data}
\label{decoding}
To decode the data, first, the cloned data is decloned from the bacterial plasmid and then it is decoded to the original file. 

First, it takes the cloned data and searches for the Cloning sites with blunt ends. After that it cuts at that ends and gets back the encoded data that was inserted in the plasmid. After decloning the data from the plasmid, it decodes the data into the original text. The Decoding algorithm is explained below: 

\begin{itemize}
    \item It takes the data and divides it into chunks of 40 bits each.
    \item Then it converts the A, T, G and C sequences back to the original binary data by using the conversion Table \ref{encodetable}. 
    \item Then it analyzes the 8-bit header and identifies which original chunks were present in it.
For instance, let $d=a \oplus b \oplus c$ be the encoded chunk formed by XOR operation on the data chunks $a$, $b$ and $c$. Suppose while decoding $b$ and $c$ are recovered and $a$ is not recovered, then  $a$ can be obtained by XORing $d$ with $b$ and $c$ ($a=d \oplus b \oplus c$ )
    \item The above step is repeated until you recover all the chunks. Then, this binary sequence is again converted back to the text data using the standard ASCII Conversion table and in this way, the original data can be obtained back.
\end{itemize}

After the decoding process, the "Gel Electrophoresis" diagram of the above experiment appears as shown in Figure \ref{capture4}. From this figure, we can see that the length of encoded data, as well as the decloned data is same, which proves the correctness of our experiment.

\section{Graphical User Interface}

\begin{figure}
\centering
\includegraphics[width=8cm]{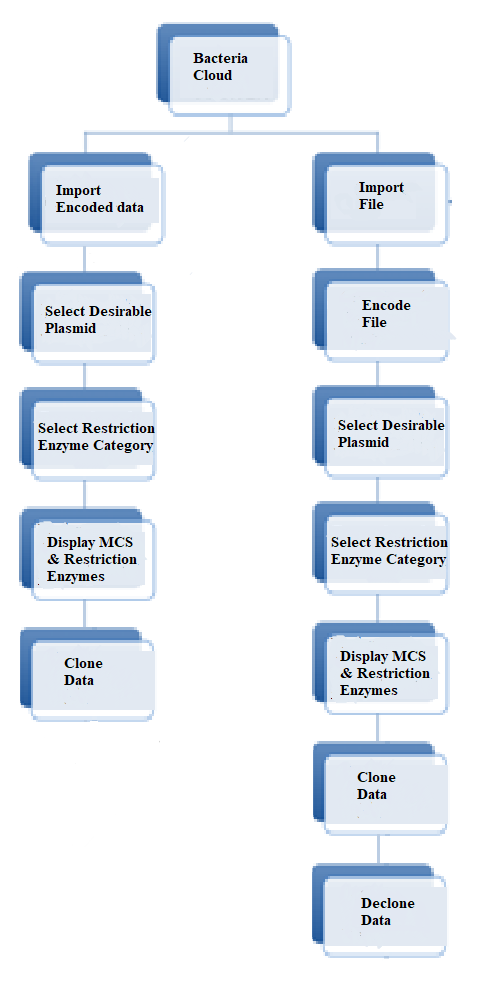}
\caption{Overview of BacSoft GUI}
\label{flowdiagram}
\end{figure}

The Graphical User Interface (GUI) of BacSoft has been developed for the users to easily upload any text file or upload any encoded DNA sequences file and get the cloned bacterial plasmid corresponding to the selected plasmid. The schematic representation of GUI is described in the Figure \ref{flowdiagram}

\subsection{Importing a text file }
\begin{figure}[H]
\centering
\includegraphics{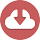}
\caption{Used to import a text file}
\label{import}
\end{figure}

As shown in the Figure \ref{import}, one can find this button on the top left corner where one can import any text file and the corresponding text data from the file can be seen on the screen under the section "Imported Data." 

\subsection{Encoding a text file}

\begin{figure}[H]
\centering
\includegraphics{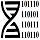}
\caption{Used for Encoding the text file}
\label{encode}
\end{figure}

As shown in the Figure \ref{encode}, by clicking this button one can convert the imported text file into corresponding DNA Encoded Sequences which can be further used for cloning purposes. Once the data is encoded, it can be seen on the screen under the section "Encoded Data." Also, the encoded data is available in the file encoded.txt at location C:/Software\_folder/plasmid/. 

\subsection{Selecting the desirable plasmid and restriction enzyme category}
\begin{figure}[H]
\centering
\includegraphics{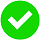}
\caption{Used after selecting Plasmid and Restriction Enzyme Category} 
\label{ok}
\end{figure}

One can find a drop-down menu beside the label "Select Desirable plasmid". There are around 176 different plasmid vectors available for options. After selecting the desirable plasmid, user can select any of the restriction enzyme categories from the drop-down menu available beside the label "Select Restriction Enzyme". After selecting the desirable plasmid and the restriction enzyme category press the "OK" button as seen in Figure \ref{ok} and the selected plasmid vector sequence is displayed on the screen under the "Plasmid" section .

\subsection{Displaying MCS and Restriction enzymes}
\begin{figure}[H]
\centering
\includegraphics{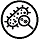}
\caption{Used for displaying MCS and Restriction Enzyme present in selected plasmid}
\label{mcs}
\end{figure}

On Clicking the button "MCS and Restriction Enzyme" as shown in Figure \ref{mcs}, two new windows pop-up appear on the screen. One of them displays the various restriction enzymes available in the selected plasmid as shown in Figure \ref{capture}, while the other one shows various antibiotic resistance markers, repressors, protein, gene, etc. available in the selected plasmid as shown in the Figure \ref{capture2}.

\subsection{Cloning data}

\begin{figure}[H]
\centering
\includegraphics{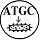}
\caption{Used for Cloning the encoded data in selected plasmid}
\label{clone}
\end{figure}

On Clicking the button "Clone Data" as shown in Figure \ref{clone}, your encoded data is inserted into the plasmid. The cloned data can be seen on the screen under the section "Cloned Data". The text in red color indicates user encoded text data while the other is the plasmid DNA sequence. A pop-up window is also displayed where the highlighted text in pink is the user encoded text data as shown in the Figure \ref{capture1}. Also, the corresponding text file named cloned\_data.txt can be found in the folder C:/Software\_folder/plasmid/. The cloning process can be visualized as shown in the Figure \ref{capture3}.

\subsection{Decloning Data}

\begin{figure}[H]
\centering
\includegraphics{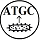}
\caption{Used for Decloning the data back from the plasmid}
\label{declone}
\end{figure}

On Clicking "Declone Data" button as shown in the Figure \ref{declone}, it retrieves the data back from the plasmid and then decodes it to the original data. Decoded data can be seen on the screen under the section "Decloned Data". In this, it also shows the "Gel Electrophoresis" experimental simulation of the Decloned data and the Encoded data as shown in the Figure \ref{capture4}. 

\section{Functionality and Workflow}
The primary objective of this subsection is to provide an overview of the working and functionality of the software.

\subsection{Importing and encoding text data}

A text file can easily be imported using the import file option. In order to clone data into the bacterial plasmid, it needs to be converted into the corresponding DNA sequences. The proposed encoding method is already described in section \ref{encoding}.

\subsection{Display MCS and Restriction Enzymes}


For cloning purpose, first, select the plasmid to clone data into it. After selecting the plasmid, for the data insertion, we have to cut the plasmid, so that the data can be inserted into it. Restriction Enzymes contains Cloning Sites from where we can cut the plasmid and insert data into it. The user can see various restriction enzymes present in the plasmid along with the Cloning sites as shown in Figure \ref{capture}.

\subsection{Clone Data}
From various restriction enzymes available, the software automatically selects restriction enzymes which will help in inserting data. After selection, the plasmid is cut from there and the Encoded data is inserted into the plasmid as shown in the Figure \ref{capture3}. In the figure, the upper portion on the right side shows the cloning sites chosen along with their restriction enzymes, while the lower right corner shows the corresponding restriction enzyme sequences.

\subsection{Declone data}
The software looks for the Restriction Enzymes with blunt ends to retrieve the data back from the plasmid,  and it is ready to be decoded. The decoding algorithm is applied as discussed in \ref{decoding}, which gives the original data back. The user can also see the "Gel Electrophoresis" simulation of the experiment as shown in Figure \ref{capture4}.



 
 \section{Examples}
 

Using BacSoft, one can import any text file and get the corresponding encoded sequences, select various restriction enzymes and MCS, clone data in the plasmid and finally visualize the Gel Electrophoresis simulation of the corresponding inserted data. Also, if the user wants to import its encoded sequence, then one can choose various restriction enzymes and MCS and the clone the data in the plasmid. The details of the examples are given below : 

\subsection{Importing the text file}  
 
\begin{figure*}
\centering
\includegraphics[scale=0.7]{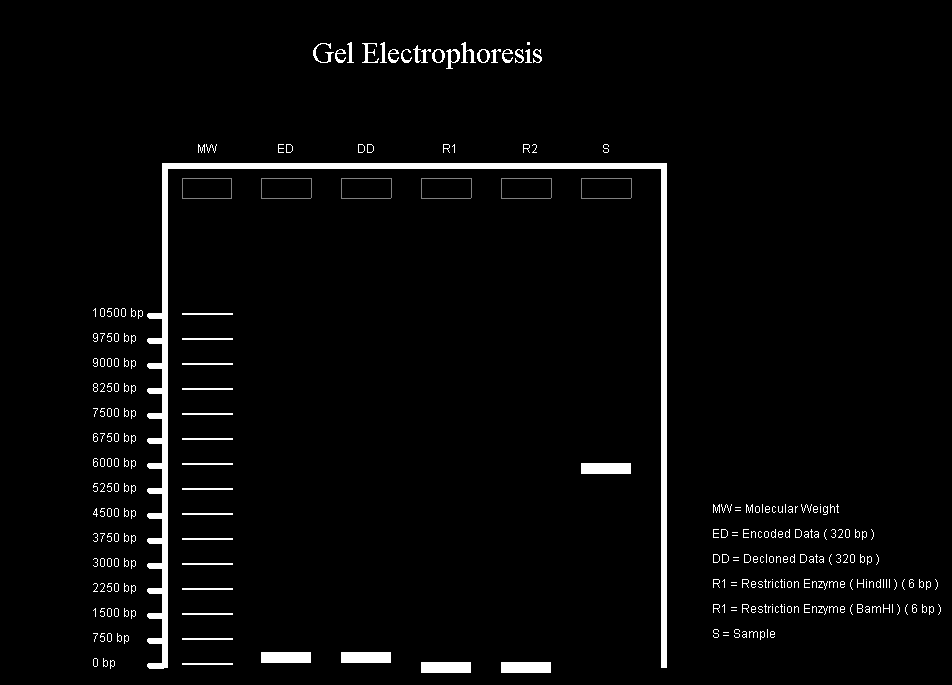}
\caption{Gel Electrophoresis simulation of the above experiment performed}
\label{capture4}
\end{figure*}

\begin{figure*}
\centering
\includegraphics[scale=0.5]{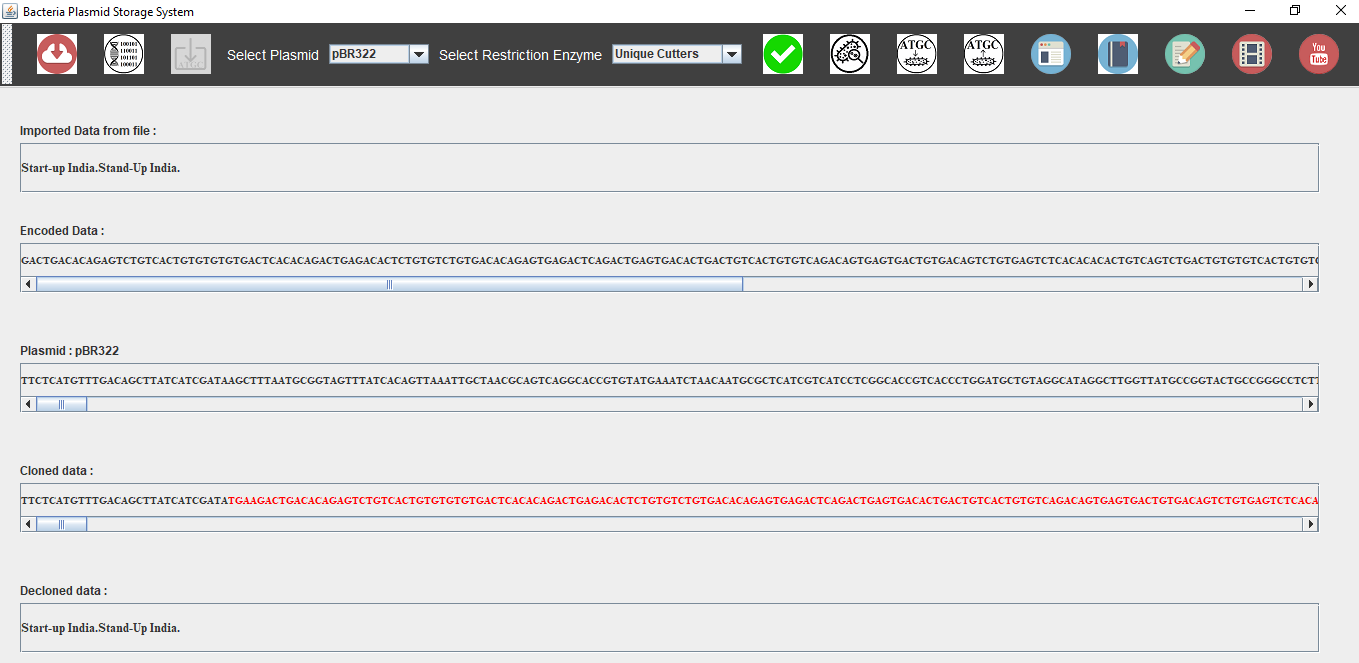}
\caption{Main Screen showing the data obtained at various steps}
\label{capture5}
\end{figure*}

' \begin{itemize}
     \item For example, let us import the text file containing the text data "Start-up India.Stand-up India." (see the Figure \ref{capture5} under the section "Imported data from file").
     
     \item Once the file is imported, the text data is encoded into the DNA sequences using the encoding strategy discussed in the section \ref{encoding}. The encoded data is of length 320 bps and can be seen in the Figure \ref{capture5} under the section "Encoded data".
     
     \item Select the plasmid "pBR322" and "Unique Cutters" restriction enzyme category for cloning the encoded data. This plasmid contains 4361 bps and 52 unique restriction enzymes. These restriction enzymes along with their cloning sites can be seen in the Figure \ref{capture}. Also, one can see various elements like antibiotic resistance markers, gene, protein, etc. present in the plasmid in the Figure \ref{capture2}.
     
     \item After this, the data is cloned into the plasmid. The encoded data is highlighted with the \textbf{pink} color in the Figure \ref{capture1}. The figure \ref{capture5} shows the cloned data under the section "Cloned data". From the Figure \ref{capture3},  it is observed that for this example, HindIII (A AGCTT) and BamHI (G GATCC) restriction enzymes are used with the Cloning sites at positions 29 and 375 bps respectively.
     
     \item After this the data is decloned back from the plasmid and decoded back using the encoding strategy discussed in  the  section \ref{decoding} to the original data as can be seen from the Figure \ref{capture5} under the section "Decloned data". Also, the Gel Electrophoresis simulation of the above example is shown in the Figure \ref{capture4}. 
     
 \end{itemize}

 \subsection{Importing DNA Sequences}
\begin{figure*}
\centering
\includegraphics[scale=0.7]{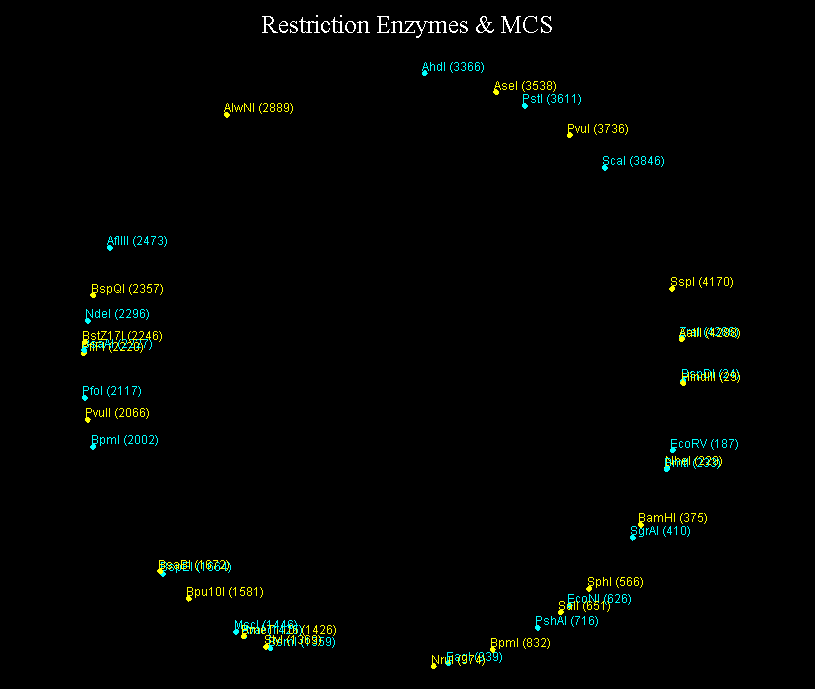}
\caption{Restriction Enzymes and MCS present in plasmid }
\label{capture}
\end{figure*}

\begin{figure*}
 \centering
\includegraphics[scale=0.7]{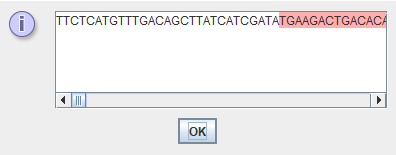}
\begin{center}
    
\end{center}
\caption{Cloned Plasmid with highlighted text representing Data}
\label{capture1}
\end{figure*}

\begin{figure*}
\centering
\vspace{-1cm}
\includegraphics[scale=0.5]{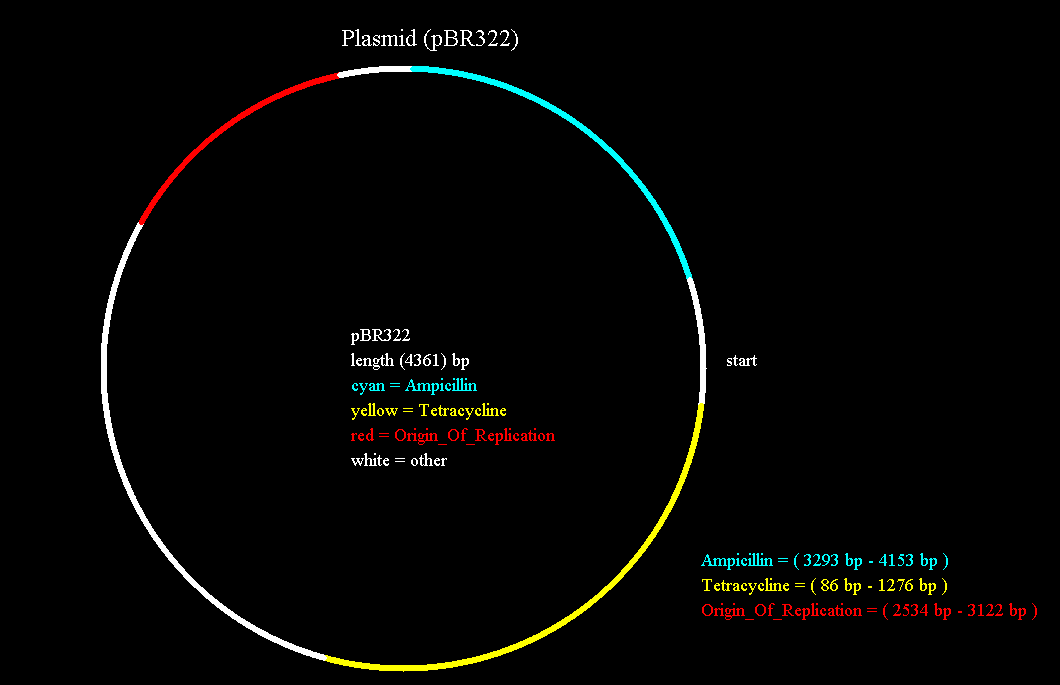}
\caption{Various elements like antibiotic resistance markers (i.e Ampicillin, tetracycline etc.), gene, protein etc. present in plasmid}
\label{capture2}
\end{figure*}

 \begin{itemize}
     \item For this example, we will import the text file containing the encoded data "AATTTTTTAAGGCC". The total length of the data encoded DNA is 14 bps.
     
    \item Select the plasmid "pBR322" and "Unique Cutters" restriction enzyme category for cloning the encoded data. This plasmid contains 4361 bps and 52 unique restriction enzymes. These restriction enzymes along with their cloning sites can be seen in the Figure \ref{capture}. Also, one can see various elements like antibiotic resistance markers, gene, protein, etc. present in the plasmid in the Figure \ref{capture2}.
     
     \item After this, the data is cloned into the plasmid. The encoded data will be highlighted in \textbf{pink} color as shown in Figure \ref{capture1}. For this example, HindIII (A AGCTT) and BsrFI (G CCGGT) restriction enzymes are used with the Cloning sites at positions 29 and 160 bps respectively.
     
 \end{itemize}
 
 
 \section{Availability}
 The BacSoft software can easily be downloaded from \href{http://www.guptalab.org/bacsoft/}{http://www.guptalab.org/bacsoft/}. 
 
 \section{Conclusion}
The software BacSoft gives a simple demonstration to encode the data in the bacterial plasmid. The software includes a data encoding method that preserves GC content and no homopolymers DNA constraints for bacterial data storage. It enables the selection of bacterial plasmid for cloning and facilitates gel electrophoresis simulation for the encoded data. An example describing the encoding and decoding of the data in the bacterial plasmid is illustrated. However, there are challenges in developing robust encoding schemes for bacterial data storage such that it achieves the maximum capacity. As a future aspect, we anticipate proposing an improved data encoding scheme by using better error correcting codes.

\bibliographystyle{plain}
\bibliography{bacterialcloud}

\end{document}